\documentclass[aps, tighten,preprint,superscriptaddress]{revtex4}
\usepackage{graphicx}
\usepackage{color}

\begin{document}

\title{Electronic and magnetic properties of V-doped anatase TiO$_{2}$
from first principles}

\author{Xiaosong Du}
\affiliation{Hefei National Laboratory for Physical Sciences at
Microscale, University of Science and Technology of China, Hefei,
Anhui 230026, P.R. China}

\author{Qunxiang Li}\thanks{Corresponding author. E-mail: liqun@ustc.edu.cn}
\affiliation{Hefei National Laboratory for Physical Sciences at
Microscale, University of Science and Technology of China, Hefei,
Anhui 230026, P.R. China}

\author{Haibin Su}
\affiliation{School of Materials Science and Engineering, Nanyang
Technological University, 50 Nanyang Avenue, 639798, Singapore}

\author{Jinlong Yang}
\affiliation{Hefei National Laboratory for Physical Sciences at
Microscale, University of Science and Technology of China, Hefei,
Anhui 230026, P.R. China}

\begin{abstract}
We report a first-principles study on the geometric, electronic and
magnetic properties of V-doped anatase TiO$_{2}$. The DFT+U (Hubbard
coefficient) approach predicts semiconductor band structures for
Ti$_{1-x}$V$_{x}$O$_{2}$ (x=6.25 and 12.5 \%), in good agreement
with the poor conductivity of samples, while the standard
calculation within generalized gradient approximation fails.
Theoretical results show that V atoms tend to stay close and result
in strong ferromagnetism through superexchange interactions. Oxygen
vacancy induced magnetic polaron could produce long-range
ferromagnetic interaction between largely separated magnetic
impurities. The experimentally observed ferromagnetism in V-doped
anatase TiO$_{2}$ at room temperature may originate from a
combination of short-range superexchange coupling and long-range
bound magnetic polaron percolation.
\end{abstract}
\pacs{62.25.+g, 77.65.-j}

\maketitle

The discovery of ferromagnetism (FM) in Co-containing TiO$_{2}$
semiconductors has attracted significant attention because of its
potential application for developing functional devices that
manipulate both spin and charge.\cite{Coexp1} However, the origin of
FM observed in Co-doped anatase TiO$_{2}$ at and above room
temperature is still under debate.\cite{review} Many experiments
argued the intrinsic nature of FM,
\cite{APL05-toyosaki,PRL05-zhao,PRL06-quilty} while the presence of
the Co clusters in the samples could not be completely
excluded.\cite{JAP03-stampe,PRB05-kim} To study the observed FM,
theoretical efforts have been made through analyzing the electronic
and magnetic structures for various Co configurations in the host
matrix. Yang \textit{et al.} have pointed out that ferromagnetic
ordering occurs only when Co atoms locate close to each
other.\cite{distribute} One plane-wave pseudopotential calculation
supports that the short-range exchange between adjacent Co atoms
could contribute to FM.\cite{janisch} In addition, oxygen vacancy is
shown to play a key role in enhancing the FM
coupling.\cite{optical,Ov1,Ov2} Recently, V-doped anatase TiO$_{2}$
has been discovered to have a surprisingly high Curie temperature
($T_{C} >$ 400 K).\cite{Vexp1} In contrast to Co-doped TiO$_{2}$
that has been widely studied theoretically, few investigations on
V-doped TiO$_{2}$ have been reported in the literature. One
interesting work by Wang \emph{et al.} using LDA and LDA+U methods
investigates the electronic structure of Ti$_{1-x}$V$_{x}$O$_{2}$
(x=6.25\%) and argues that the neighboring V-V cations prefers FM at
a high doping concentration (x=25\%) by employing a
2$\times$1$\times$1 supercell.\cite{Vcal} However, the relative
stability among magnetic phases is extremely sensitive to atomistic
structure, its associated defects' distribution and dopant
concentration, further investigation is desirable to gain better
understanding of high $T_{C}$ FM in this system.

In this paper, we use \textit{ab initio} band-structure and total
energy methods, implemented in the Vienna \textit{Ab initio}
Simulation Package (VASP),\cite{paw,vasp} to study the geometric,
electronic and magnetic properties of V-doped anatase TiO$_{2}$ (
Ti$_{1-x}$V$_{x}$O$_{2}$), in particular the effects of doping
concentration (x=6.25\% and 12.5\%) as well as oxygen vacancy. The
projector augmented wave (PAW) method is chosen to represent the
electron-ion interaction.\cite{paw} Exchange correlation
interactions are described by the Perdew-Burke-Ernzerhof generalized
gradient approximation (GGA).\cite{pbe} The Brillouin-zone (BZ)
integration is performed on a well converged Monkhost-Pack k-points
grid\cite{mp}. The plane wave kinetic energy cutoff is set to be 400
eV. Atomic positions and lattice parameters are optimized at GGA
level until the atomic forces are smaller than 0.01 eV/\AA. To
account for the strongly correlated interactions of the V
3\textit{d} electrons, a moderate on-site Coulomb repulsion (U=3.0
eV) has been applied only to V 3\textit{d} orbitals since further
correction for the host material has little impact on the magnetic
ordering.\cite{ZnO}

We start with a Ti$_{15}$VO$_{32}$ supercell to model the low doping
concentration case (6.25\%), where a 48-atom 2$\times$2$\times$1
supercell of pure anatase TiO$_{2}$ (Ti$_{16}$O$_{32}$) is
demonstrated in Fig.1 and one V atom substitutes Ti at Ti$_{1}$
site. The lattice parameters of optimized Ti$_{15}$VO$_{32}$
supercell are a=7.64 {\AA} and c=9.65 \AA, which are slightly
smaller than those of perfect TiO$_{2}$ (our theoretical lattice
parameters of Ti$_{16}$O$_{32}$ are a=7.65 {\AA} and c=9.68 \AA
\space at GGA level). Clearly, this type of substitution only leads
to a small geometrical distortion in the vicinity of V impurity
(less than 0.1 \AA). To discuss the solubility of V in anatase
TiO$_{2}$ host, it is useful to define the substitution energy of V
impurity as
E$_{s}$=E(Ti$_{15}$VO$_{32}$)+E(Ti)-E(V)-E(Ti$_{16}$O$_{32}$), where
E(Ti$_{16}$O$_{32}$),E(Ti) and E(V) is the total energy of a
2$\times$2$\times$1 supercell of pure anatase TiO$_{2}$, the bulk
hcp Ti and bcc V, respectively. $E_{s}$ is predicted to be 2.34 eV,
which is even less than one third of the substitution energy of Co
dopant (8.51 eV).\cite{distribute} This result agrees nicely with
experimental observation that V impurity is well dissolved into
TiO$_{2}$.\cite{Vexp1}

The band structure, total density of states (DOS) and V 3\textit{d}
partial DOS (PDOS) of Ti$_{15}$VO$_{32}$ are presented in
Fig.\ref{low}. The calculation at GGA level suggests that the DOS
near the Fermi surface ($E_{F}$) mainly originates from V $t_{2g}$
orbitals, which is shown in Fig.\ref{low}(c) for clarity. The V
impurity band has strong anisotropic character in the $a$-$b$ plane
(i.e. A$\rightarrow$R and $\Gamma$$\rightarrow$$M$), however, it is
less dispersive along the $c$ axis of the BZ (\textit{i.e.}
M$\rightarrow$A and Z$\rightarrow$$\Gamma$). It is important to note
that $E_{F}$ is largely crossed by spin-up states, while spin-down
states only cut $E_{F}$ slightly. This indicates that the system is
nearly half-metallic at GGA level. As there exists strongly
correlated interaction in V 3\textit{d} shells, we choose DFT+U as a
scheme beyond-GGA level. The calculated PDOS by DFT+U are plotted in
Figs.2(e) and 2(f). We find that the metal-insulator-transition
occurs when the parameter U is applied to the V atom. This strong
correlation lifts the degeneracy of $t_{2g}$ states, so as to open
the gap. An occupied band composed of mainly $d_{xy}$ component is
split off from the bottom of the conduction band. These features
suggest that Hubbard $U$ correction increases the localization of
the V $d$ electrons. Results presented above are fundamentally
consistent with Ref.\cite{Vcal} except for their lifted Ti PDOS due
to additional corrections for Ti 3\textit{d} states.

It can be expected that the direct impurity interaction can not
yield experimentally observed strong FM because of the large V-V
distance. The original supercell is extended along $b$ and $c$ axis
to form 2$\times$4$\times1$ and 2$\times$2$\times2$ supercells
(Ti$_{30}$V$_{2}$O$_{64}$) to examine magnetic ordering. From Table
I, the calculated energy difference between the AFM and FM states
($\Delta$E) is negative, which indicates that the ground state is
AFM for both cases as expected. Moreover, the on-site Coulomb
repulsion at V atom strengthens AFM with a small increase of
magnetic moment. The AFM ground state is further confirmed by our
calculation using hybrid B3LYP functional as implemented in
CRYSTAL03 code.\cite{b3lyp} To understand the experimental observed
high $T_{C}$ in V-doped TiO$_{2}$, one has to search other origins
to lead FM. In the rest of this paper, first we explore the effects
of closely distributed V impurities esp. at high doping level, then
oxygen vacancy on electronic and magnetic properties.

According to Janisch \textit{et al.}'s
suggestion,\cite{review,janisch} we have constructed three
Ti$_{14}$V$_{2}$O$_{32}$ structures to model high impurity
concentration (12.5\%) of V-TiO$_{2}$, where a second V atom
occupies Ti$_{2}$, Ti$_{3}$ or Ti$_{4}$ site in the
Ti$_{15}$VO$_{32}$ supercell, named the chain, grid and
nearest-neighbor (NN) configurations (the optimized V-V distance is
3.82, 5.39 and 3.06 \AA), respectively. From Table I, $\Delta$E is
positive for all three cases within GGA approach and increases with
V-V distance decreasing. We find that the energy of the NN
configuration is the lowest, strongly favoring V-dopants clustering.

The DOS shapes have almost the same characteristic shapes among
three configurations, hence, only the one for the nearest-neighbor
one is presented in Fig.\ref{nn}. Similar to the Co-TiO$_{2}$
case,\cite{janisch} when $x$ increases from 6.25\% to 12.5\%, the
total DOS, V 3\textit{d} PDOS and magnetic moment of V cation change
slightly. The DOS and PDOS by DFT+U are presented in Figs.3(b) and
3(d), which reveal that the system becomes semiconductor when the
parameter U is applied. It is consistent with the poor conductivity
observed in V-TiO$_{2}$ samples. The on-site U correction shifts
$d_{xy}$ state down to the middle gap, while at GGA level it is a
mixture of three $t_{2g}$ orbitals ($d_{xy}$, $d_{yz}$ and $d_{xz}$)
locating at $E_{F}$. The larger separation between the occupied and
empty V 3\emph{d} states, due to strong correlation, results in the
decrease of the $\Delta$E value and even the change of its sign for
the chain and grid configurations in Table I . The remarkable
observation in our work is the high fidelity of $\Delta$E for the NN
configuration. As it is extremely hard to choose the accurate value
for U, this result is particularly valuable to highlight the
important contribution of V-O-V NN configuration to yield FM state.

The reason for the weakened FM couplings within DFT+U could be
sought in terms of the $p$-$d$ hopping mechanism\cite{ZnO}. The V
impurities interact with orbitals of the same symmetry and form a
set of bonding-antibonding states. Within GGA approximation, FM
state has an energy gain through the partial occupancy of V $t_{2g}$
states. The Hubbard U at V sites results in a fully occupied
$d_{xy}$ band. In a FM arrangement, both bonding and antibonding
levels are filled, and there is no energy gain for this coupling. In
the AFM arrangement, the bonding states are filled for both spin up
and down channels, while the antibonding ones are empty. It implies
that AFM is favored between adjacent V cations when correlation
effect is taken into account. However, there exists a strong
competition due to superexchange. It is well-known that
superexchange may have a FM contribution for filled shells.
Recently, Janisch \textit{et al.} argue that short-range
superexchange between 90$^{\circ}$ metal-anion-metal bond plays a
significant role in transition metal doped anatase
TiO$_{2}$.\cite{janisch} We believe the FM originates from the
superexchange interaction with the V-O-V bond angle about
104.9$^{\circ}$ in the NN configuration.

As indicated by previous work\cite{Ov1,Ov2}, the presence of oxygen
vacancy ($V_{O}$) may influence the distribution of dopants. To
study the impact of $V_{O}$, three cases are considered: (a)
Ti$_{16}$O$_{31}$ with one oxygen atom removing from the
2$\times$2$\times$1 supercell of TiO$_{2}$; (b) The basal oxygen
(O$_{1}$) of V-contained octahedron is removed from
Ti$_{15}$VO$_{32}$ and subsequently the oxygen deficient cell is
doubly enlarged along $b$ axis, named as the basal O$_{1}$
2$\times$4$\times$1 configuration; (c) $V_{O}$ occupies the vertex
site of V-incorporated octahedron (by removing O$_{2}$) and the
original supercell is doubly extended along $c$ axis, denoted as the
vertex O$_{2}$ 2$\times$2$\times$2 configuration. After removing an
oxygen atom from the cation-contained octahedron, the original
structure experiences a considerable distortion. Cations next to the
$V_{O}$ site are repelled away. The calculated total energies
indicate that $V_{O}$ prefers to stay close to V than near Ti.

Fig.\ref{basal} shows the DOS results for the Ti$_{16}$O$_{31}$ and
basal O$_{1}$ 2$\times$4$\times$1 configurations. When an oxygen
atom is removed, a nonmagnetic defect state (mainly from Ti cations
near $V_{O}$) appears at the bottom of the conduction band. If a V
atom  occupies the cation site of the oxygen-deficient octahedron, V
3\textit{d} states overlap in energy with the defect state which
yields 0.11 e transferred from the defect state to the empty V
3\textit{d} spin-up state enhancing spin polarization of V ion. It
fits quite well with the physical picture proposed by Coey
\textit{et al.} that the hybridization between the defect state and
magnetic dopants' states will promote FM.\cite{Coey} However, the
$V_{O}$ mediated magnetic coupling is sensitive to
distance.\cite{polaron} The polarons induced by $V_{O}$ tend to
parallelize their spins, corresponding to a ferromagnetic
arrangement, only when their distance is less than a critical value,
which is the case in the basal O$_{1}$ 2$\times$4$\times$1
configuration. When their distance increases larger than this
critical value, the strict spin alignment between polarons
disappeared. This leads to a loss of long-range coherence, as in the
vertex O$_{2}$ 2$\times$2$\times$2 case. Further efforts are needed
to determine the exact critical distance required for polaron
percolation in this system. DFT+U calculations qualitatively do not
change the result given by GGA.

In summary, we have studied the structural, electronic and magnetic
properties of V-doped anatase TiO$_{2}$ by both GGA and DFT+U
calculations. The correlation effect is not well represented by GGA,
while DFT+U calculations give a better physical picture. The results
suggest that direct exchange between well separated V ions cannot
account for the observed FM in this system. The superexchange
interaction is responsible for the stable FM between
nearest-neighbor V cations. The long-range interactions between
oxygen defects induced magnetic polarons are sensitive to their
distance. Provided that a macroscopic ferromagnetic clusters appear
by polarons overlapping, the system favors FM. Thus, a combination
of the long-range polaron percolation and the short-range
superexchange through nearest-neighbor V atoms could contribute to
the high $T_{C}$ FM in V-doped anatase TiO$_{2}$.

We thank Y.Wang for helpful discussion. This work is partially
supported by the National Natural Science Foundation of China under
Grand Nos. 20303015, 10674121, 50121202 and 20533030, by the USTC-HP
HPC project, and by the SCCAS and Shanghai Supercomputer Center.

\newpage
\begin{table}[htbp]
\caption{Calculated energy difference between the AFM and FM states
($\Delta$E=E(AFM)-E(FM)), and the ground state magnetic moment (M)
per V atom for various configurations at GGA level and at DFT+U
level. The data inside the parentheses are obtained by DFT+U
approach.} \label{table1} \begin{tabular}{lrr} \hline Unit cells &
$\Delta$E(meV/atom) & M($\mu_{B}$) \tabularnewline \hline
Ti$_{30}$V$_{2}$O$_{64}$: & & \tabularnewline $2\times2\times2$ &
-1.0 (-4.2) & $_{-}^{+}$0.71 ($_{-}^{+}$0.83) \tabularnewline
$2\times4\times1$ & -8.0 (-12.7) & $_{-}^{+}$0.69
($_{-}^{+}$0.80)\tabularnewline Ti$_{14}$V$_{2}$O$_{32}$: & &
\tabularnewline chain & 60.2 (-5.0) & 0.84 ($_{-}^{+}$0.81)
\tabularnewline grid & 21.8 (-56.0)& 0.77 ($_{-}^{+}$0.83)
\tabularnewline nearest-neighbor & 75.2 (14.5) & 0.89
(0.91)\tabularnewline Ti$_{30}$V$_{2}$O$_{62}$: & & \tabularnewline
basal O$_{1}$ $2\times4\times1$ & 18.5 (18.5)& 1.14
(1.37)\tabularnewline vertex O$_{2}$ $2\times2\times2$ & -3.8 (-4.5)
& $_{-}^{+}$0.97 ($_{-}^{+}$1.17) \tabularnewline \hline
\end{tabular}
\end{table}

\newpage
\begin{figure}[!htp]
\includegraphics[width=7cm]{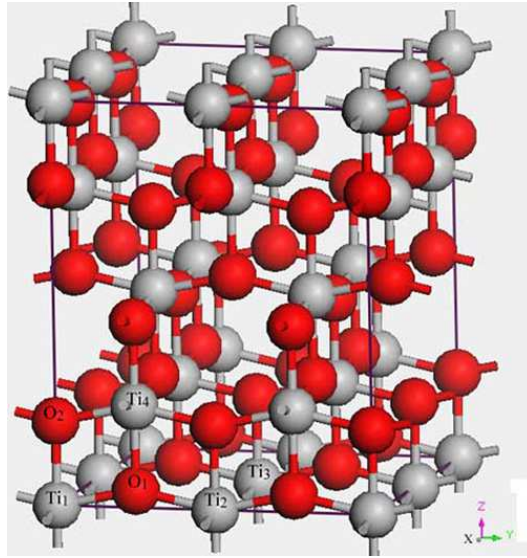}
\caption{(Color online) The schematic 48-atom supercell of anatase
TiO$_{2}$ (2$\times$2$\times$1). The gray and red balls stand for Ti
and O atoms, respectively.} \label{Ti}
\end{figure}

\newpage
\begin{figure}[!htp]
\includegraphics[width=7.5cm]{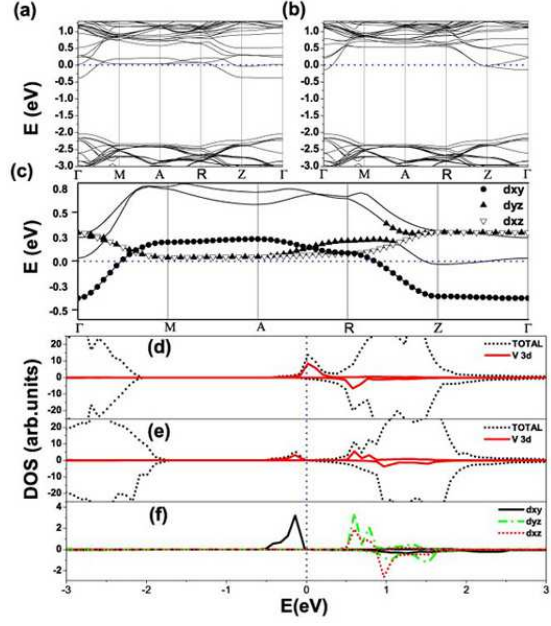}
\caption{(Color online) Band and DOS structure of V-doped anatase
TiO$_{2}$ modeled by a $2\times2\times1$ supercell
(Ti$_{15}$VO$_{32}$). (a),(b) the majority and minority spin GGA
band structure. (c) majority spin around $E_{F}$. Total DOS and V
3\textit{d} PDOS (d) GGA results, (e) DFT+U results. (f) DFT+U
results of the projected DOS of V 3\textit{d} orbitals. The Fermi
energy is shifted to zero. $\Gamma$=(0,0,0), M=(1/2,1/2,0),
A=(1/2,1/2,1/2), R=(1/2,0,1/2), and Z=(0,0,1/2).} \label{low}
\end{figure}

\newpage
\begin{figure}[!htp]
\includegraphics[width=7cm]{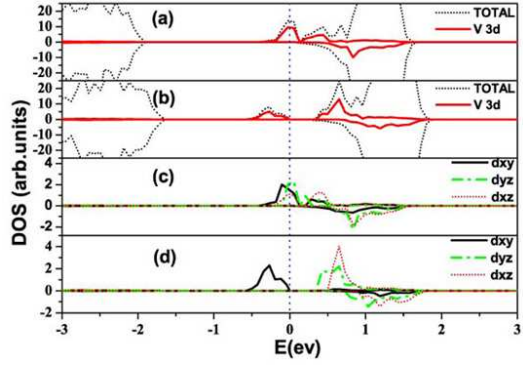}
\caption{(Color online) DOS structure of the nearest-neighbor
configuration (Ti$_{14}$V$_{2}$O$_{32}$) with a ferromagnetic
arrangement of the V moments. Total DOS and V 3\textit{d} PDOS (a)
GGA result, (b) DFT+U result. Projected DOS of V 3\textit{d}
orbitals (c) GGA result, (d) DFT+U result. The vertical line denotes
the Fermi level.} \label{nn}
\end{figure}

\newpage
\begin{figure}[!htp]
\includegraphics[width=7cm]{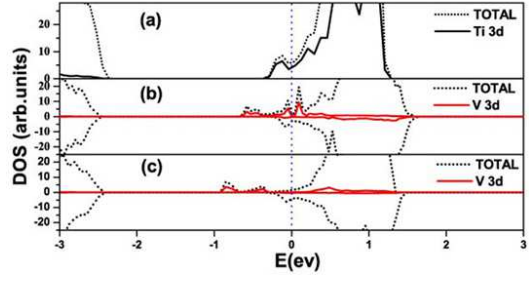}
\caption{(Color online) DOS structure of oxygen deficient
Ti$_{16}$O$_{31}$ and the basal O$_{1}$ $2\times4\times1$
configuration (Ti$_{30}$V$_{2}$O$_{62}$). (a) Total DOS and
3\textit{d} PDOS of Ti cations near the oxygen defect for
Ti$_{16}$O$_{31}$. Total DOS and V 3\textit{d} PDOS for the basal
O$_{1}$ $2\times4\times1$ obtained within (b) GGA approach, (c)
DFT+U method. The vertical line denotes the Fermi level.}
\label{basal}
\end{figure}

\end{document}